\documentclass[jcp,article,submit,pdftex,moreauthors]{Definitions/mdpi} 

\firstpage{1} 
\pubvolume{1}
\issuenum{1}
\articlenumber{0}
\pubyear{2023}
\copyrightyear{2023}
\datereceived{ } 
\daterevised{ } %
\dateaccepted{ } 
\datepublished{ } 
\hreflink{https://doi.org/} %

\usepackage{amsmath,amssymb,amsfonts}
\usepackage{algorithmic}
\usepackage[T1]{fontenc}
\usepackage{changes}
\usepackage{graphicx}
\usepackage{listings}
\usepackage{tabularx}
\usepackage{textcomp}
\usepackage{subfig}
\usepackage[normalem]{ulem}

\newcolumntype{Y}{>{\centering\arraybackslash}X}
\renewcommand{\linenumbers}{}

\Title{Towards a Near-real-time Protocol Tunneling Detector based on Machine Learning Techniques}

\TitleCitation{Towards a Near-real-time Protocol Tunneling Detector based on Machine Learning Techniques}

\Author{Filippo Sobrero $^{1}$, Beatrice Clavarezza $^{1}$, Daniele Ucci $^{1}$, and Federica Bisio $^{1}$}

\AuthorNames{Filippo Sobrero, Beatrice Clavarezza, Daniele Ucci and Federica Bisio}

\AuthorCitation{Sobrero, F.; Clavarezza, B.; Ucci, D.; Bisio, F.}

\address{%
$^{1}$ \quad aizoOn Technology Consulting; name.surname@aizoongroup.com}

\conference{IEEE Symposium Series on Computational Intelligence, 2021}

\abstract{In the very last years, cybersecurity attacks have increased at an unprecedented pace, becoming ever more sophisticated and costly. Their impact has involved both private/public companies and critical infrastructures.
At the same time, due to the COVID-19 pandemic, the security perimeters of many organizations expanded, causing an increase of the attack surface exploitable by threat actors through malware and phishing attacks.
Given these factors, it is of primary importance to monitor the security perimeter and the events occurring in the monitored network, according to a tested security strategy of detection and response.
In this paper, we present a protocol tunneling detector prototype which inspects, in near real time, a company's network traffic using machine learning techniques.
Indeed, tunneling attacks allow malicious actors to maximize the time in which their activity remains undetected.
The detector monitors unencrypted network flows and extracts features to detect possible occurring attacks and anomalies, by combining machine learning and deep learning.
The proposed module can be embedded in any network security monitoring platform able to provide network flow information along with its metadata.
The detection capabilities of the implemented prototype have been tested both on benign and malicious datasets.
Results show $97.1\%$ overall accuracy and an F$1$-score equals to $95.6\%$.}
\keyword{passive network analysis; dns tunneling; anomaly detection; machine learning; deep learning} 

\begin{document}

\section{Introduction}
\label{sec:intro}

Cybersecurity attacks keep increasing year over year at an unprecedented pace, becoming ever more sophisticated and costly~\cite{ETL2022, IBM2022}.
The growth between 2021 and 2022 has resulted in a rise of attacks' volume and impact on both private/public companies and critical infrastructures.
Companies comprise digital service providers, public administrations and governments, and include businesses operating in finance and health sectors.
In particular, service providers have experimented more than 15\% raise in intrusions (infamous has been the case of Solarwinds~\cite{Solarwinds2020}) compared to 2021~\cite{ETL2022}, a trend destined to grow in the next years~\cite{Gartner2022}.
At the same time, due to the COVID-19 pandemic, the security perimeters of many organizations expanded to cope with the new needs of remote working, causing an increase of the attack surface exploitable by attackers~\cite{Gartner2022}.
The European Union Agency for Cybersecurity estimates that more than $10$ terabytes of data are stolen monthly from target assets that are made unavailable, until a ransom is payed~\cite{ETL2022}, while IBM calculates that the average cost of these attacks is $\$4.54$M, arriving up to $\$5.12$M~\cite{IBM2022}.
On the other hand, malware attacks are still on the rise after the pause recorded during the pandemic and phishing continues to be the common attack vector for initial access~\cite{ETL2022}.

Given these factors, it is of primary importance to monitor the security perimeter and the events occurring in the network, according to a tested security strategy of detection and response.
According to Gartner~\cite{Gartner2022}, newly proposed solutions should be automated as much as possible, since human errors continue to play a crucial role in most security breaches.
In this paper, we present a protocol tunneling detector prototype which inspects -- in near real time -- a company's network traffic using machine learning.
Tunneling techniques allow attackers to create a tunnel through a network by encapsulating traffic inside another protocol~\cite{TunnelMITRE2020} and, hence, can be used to let infected machines to contact their corresponding command-and-control centers.
Thus by abusing legitimate network traffic protocols, like DNS~\cite{ENISAETA2020}, the attacker maximizes the time in which the infection remains undetected.
In this work, we rely on a commercial network security monitoring platform for detecting and investigating potentially malicious or anomalous activities~\cite{bisio2017real,lombardo2018fast,Saeli2020,DBLP:conf/ssci/UcciSBZ21}, but the proposed solution can be easily integrated into any network security monitoring platform able to provide network flow information along with its metadata. 
The platform we employ is responsible for collecting, processing network flows and dispatching them to one, or more, advanced cybersecurity analytics (ACAs), which are able of recognizing the signals of possible occurring attacks and anomalies.
In this scenario, the detector monitors only cleartext protocols, but it works jointly with an ACA responsible for analyzing encrypted traffic~\cite{DBLP:conf/ssci/UcciSBZ21}. 
Indeed, while some cleartext protocols are extensively used (i.e., DNS), nowadays the vast majority of Internet traffic is encrypted~\cite{Felt2017,VMware2020,Decipher2019,Fortinet2020,GoogleChrome2020}: this enabled threat actors to perform malware campaigns relying on HTTPS for delivering malware and contacting command-and-control centers~\cite{Cisco2019}.
Just in 2020, $67\%$ of malware has been delivered via encrypted HTTPS connections~\cite{ENISA2020}.
Along with malware delivery, malicious secure communications are used to exfiltrate data and steal sensitive information from private and public companies~\cite{korolov2012cyber,taylor2014cyber,yadav2016cyber_sec}.
While the analytics dealing with encrypted traffic has been extensively described in~\cite{DBLP:conf/ssci/UcciSBZ21}, we extend this previous work by backing up secure connection analysis to the monitoring of cleartext protocols.
As mentioned before, the latter can be used to discover the abuse of and signal network packets' contents which are not usually observed in the monitored network.
The module, presented in this paper, extracts a sequence of \textit{N} bytes of each single network packet and computes features associated to the collected stream of bytes.
Through the combination of deep learning and machine learning, each network packet is assigned to a specific network protocol and, if a connection exhibits anomalies (e.g., an interleaving of different protocols), a security analyst is notified about the discovered inconsistency.
More specifically:
\begin{itemize}
    \item we implement a protocol tunneling detector prototype which analyzes, in near real time, a byte sequence of the packets flowing in the monitored network
    \item the proposed prototype combines
    \begin{itemize}
        \item an artificial neural network (ANN), based on~\cite{SANS2021}, that accurately classifies cleartext protocols and identify possible anomalies in network connections
        \item a support vector machine that is able to detect compressed/encrypted traffic within unencrypted connections
    \end{itemize} 
    \item we design and implement an input sanitization module, which automatically removes inconsistent data from models' training sets to significantly increase the models' performance
\end{itemize}
With respect to~\cite{SANS2021}, we changed both the input byte sequences we provide to the ANN and their sizes in bytes (as detailed in Sections~\ref{sec:approach} and~\ref{sec:eval}).
The performance of the proposed approach has been evaluated on different datasets that either contain legitimate traffic or simulate DNS tunneling attacks, which are the most common~\cite{ENISAETA2020}.
The obtained overall accuracy of the proposed prototype is $97.1\%$, along with an F$1$-score equals to $95.6\%$.

The rest of the paper is organized as follows: Section~\ref{sec:related} discusses related work, while Section~\ref{sec:back} introduces basic notions that will be later used to detail the proposed approach (Section~\ref{sec:analytics}).
The experimental evaluation is reported in Section~\ref{sec:eval} and, finally, Section~\ref{sec:conclusion} concludes the paper.
\section{Related Work}
\label{sec:related}

Tunneling attacks are a specific typology of network attacks in which an attacker creates a tunnel through a network by encapsulating traffic inside another protocol~\cite{TunnelMITRE2020}.
This allows the attacker to bypass traditional network security controls and potentially exfiltrate sensitive information.
Therefore, as discussed in Section~\ref{sec:intro}, using cleartext network protocols may pose a significant risk when these are abused by malicious actors.

In this context, DNS tunneling represents one of the most common techniques employed for covertly exfiltrating data from a network, by encoding the data in DNS queries and responses. Since this method is becoming increasingly prevalent, a growing body of research aims at detecting and mitigating DNS tunneling attacks. In~\cite{wang2021comprehensive}, the authors review detection technologies from a perspective of rule-based and model-based methods with descriptions and analyses of DNS-based tools and their corresponding features, covering detection approaches developed from 2006 to 2020 by means of a comparative analysis.

Latest works in the area of DNS tunneling detection mainly cover three main categories, i.e., detection approches via machine learning, real-time detection approaches, and detection of DNS tunneling variants (e.g., fast flux~\cite{lombardo2018fast}, and domain generation algorithms (DGAs)~\cite{bisio2017real}).

Regarding the first group, researchers have recently proposed deep learning algorithms such as Convolutional Neural Networks (CNNs) and Recurrent Neural Networks (RNNs) for detecting DNS tunneling traffic. In \cite{palau2020dns}, the authors develop a novel DNS tunneling detection method employing a Convolutional Neural Network (CNN) to analyze DNS queries and responses and identify DNS tunneling activities. The proposed approach is evaluated using a dataset of real-world DNS traffic and show promising results in detecting DNS tunneling attacks with high accuracy. The work of \cite{zhang2019dns} apply both Convolutional Neural Networks (CNNs) and Recurrent Neural Networks (RNNs) for detecting DNS tunneling traffic. The authors show that these algorithms can effectively spot and identify malicious patterns.

The second group of studies has focused on developing real-time detection systems for DNS tunneling.
These systems use a combination of several detection techniques to quickly identify malicious DNS traffic~\cite{ahmed2019real}.
In~\cite{rajendran2020dns}, the authors present an overview of principal countermeasures for DNS tunneling attacks.

Regarding the state of the art of approaches that analyze encrypted communications, it has already been presented in~\cite{DBLP:conf/ssci/UcciSBZ21}.
The approach we present and evaluate in the next sections passively extracts both sequential and statistical features from network flows to detect tunneling attacks in cleartext protocols.
As sequential features, we refer to those characteristics obtained from raw flow sequences.
Differently from the works previously described in this section, we directly examines, for each packet, a specific sequence of bytes for tunneling detection by using artificial neural network, which are simpler deep learning models.

\section{Background}
\label{sec:back}

\subsection{DNS Tunneling}
\label{sec:dns_tunneling}
\begin{figure}[tp]
    \centering
    \includegraphics[width=0.9\textwidth]{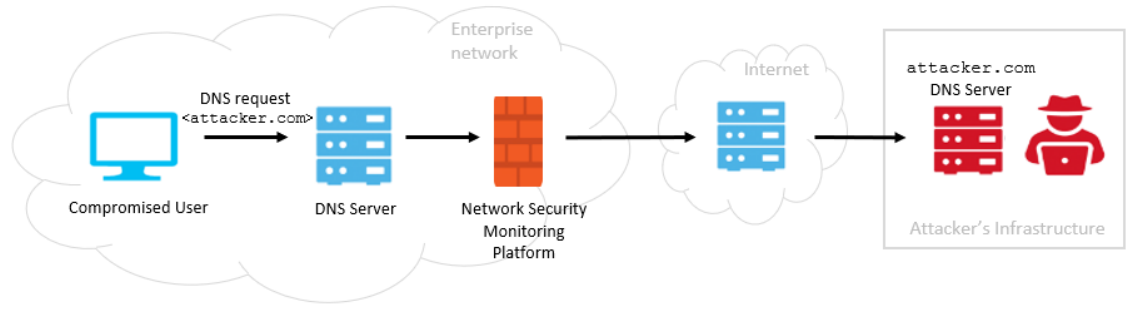}
    \caption{A DNS tunneling example}
    \label{fig:dns_tunnel}
\end{figure}

Protocol tunneling is an attack technique commonly used to maximize the time in which the infection remains undetected in a targeted network.
In this context, the DNS protocol is usually abused in order to bypass security gateways and, then, to tunnel malware and other data through a client-server model~\cite{ENISAETA2020}.
Figure~\ref{fig:dns_tunnel} depicts a typical DNS tunneling scenario: firstly, an attacker registers a malicious domain (e.g., \texttt{attacker.com}) on a C\&C center managed by her; at that point, assuming that the attacker has already taken control over a machine inside the targeted network and violated its security perimeter, the infected computer sends a query to the malicious domain.
Since DNS requests are typically allowed to move in and out of the network, the query through the DNS resolver reaches the attacker’s C\&C center, where the tunneling program is installed.
This established tunnel can be used either to exfiltrate data and sensitive information or for other malicious purposes.

\subsection{Support Vector Machines}
\label{sec:one_class}
The original formulation of Support Vector Machines~\cite{vapnik2013nature} (SVMs) is related to the resolution of supervised tasks with the objective of finding a maximum margin hyperplane that separates two or more classes of observations.
In the last years, also one-class SVMs have been shown to represent a suitable choice in the context of anomaly detection~\cite{swersky2016evaluation}. 
It is defined as a boundary-based anomaly detection method, which modifies the original SVM approach by extending it in order to deal with unlabeled data.
Like traditional SVMs, one-class SVMs can also benefit of the so called kernel trick when extended to non-linearly transformed spaces, by defining an appropriate scalar product in the feature space.

\subsection{Artificial Neural Networks}
\label{sec:ann}

Artificial Neural Networks (ANNs) are deep learning models that have been successfully applied to a vast number of knowledge fields ranging from computing science to arts~\cite{abiodun2018state}.
They are internally constituted by groups of multiple neurons, which can be thought of as mathematical functions that take one or more inputs.
In ANNs, inputs are processed only forward and are multiplied by weights within each neuron and summed up to be then passed to an activation function and become the neuron’s output.
In general, artificial neural networks consist of three different layers: input, hidden and output; the first layer accepts inputs, while the hidden layers process them to learn the optimum weights.
Finally, the output layer produces the result.
\section{Protocol tunneling detector}
\label{sec:analytics}

The proposed architecture splits the burden of processing the traffic of a monitored network into two different sub-modules: the first mainly deals with secure connections, while the second inspects unencrypted traffic.
As previously discussed, the former analytics has been detailed in~\cite{DBLP:conf/ssci/UcciSBZ21}.
At a glance, it detects possible anomalies occurring during a SSL/TLS handshake between a client, located inside the network monitored by the software platform outlined in Section~\ref{sec:intro}, and an external server. 
The SSL/TLS detection analytics examines information contained in X.509, SSL, and TLS exchanged protocol messages.
Instead, the second module looks for anomalies in unencrypted traffic, regarding the abuse of specific protocols (i.e., tunneling attack techniques).
To provide these detection capabilities, this prototype collects a sequence of bytes from each network packet and inspects its content.
The content, along with its features, is fed to a testing module, which detects possible anomalies that are signaled to security analysts.

\subsection{General approach}
\label{sec:approach}

\begin{figure}[tp]
	\centering
	\includegraphics[width=\textwidth]{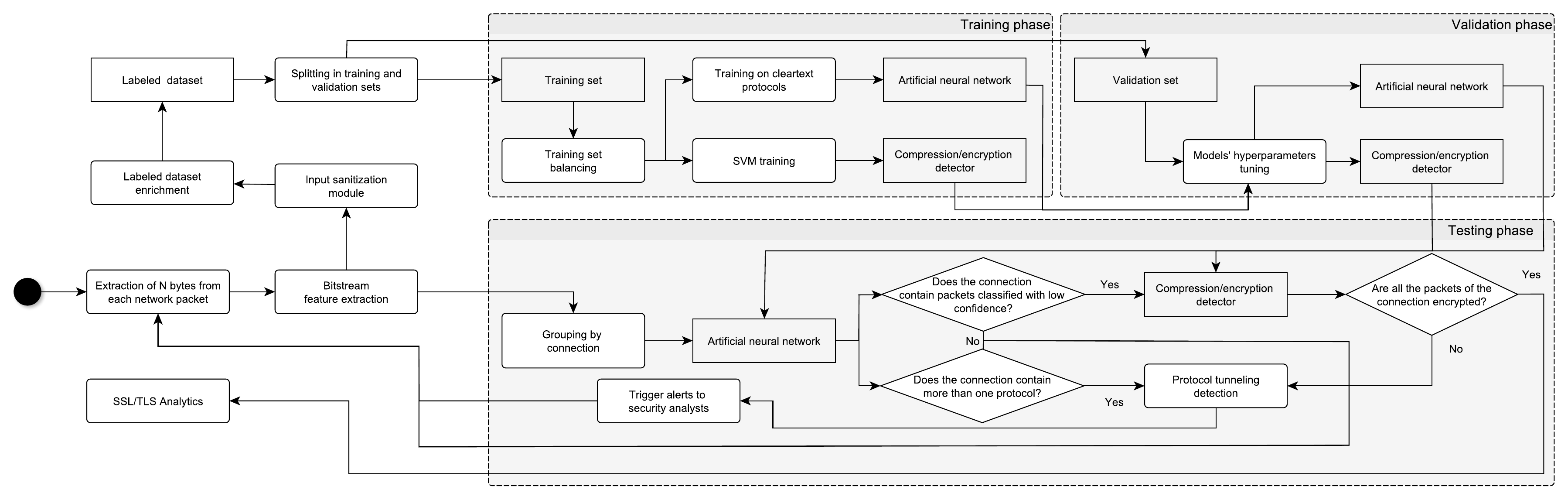}
	\caption{Protocol tunneling detector prototype overview.}
	\label{fig:proto_flow}
\end{figure}

Figure~\ref{fig:proto_flow} reports the general structure of the proposed anomaly detection methodology: for each packet observed in the network, the prototype collects a sequence of \textit{N} bytes belonging to the highest network protocol used in the communication.
As an example, in a secure connection which relies on HTTPS, the bytes returned by the extraction process are the ones related to HTTPS, and not to the other lower-layer protocols (e.g., TCP).
\begin{figure}[bp]
	\centering	\includegraphics[width=\textwidth]{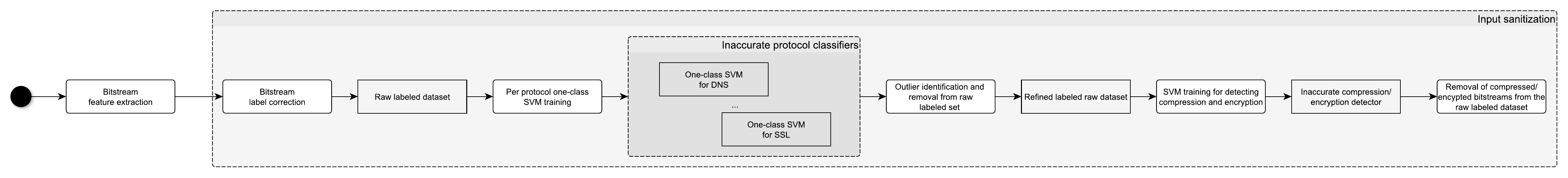}
	\caption{Input sanitization module.}
	\label{fig:san_mod}
\end{figure}
From the obtained bitstream, we extract the following sequential features (i.e., those features obtained from raw flow sequences):
\begin{itemize}
    \item binary representation of collected bytes
    \item bitstream entropy and \textit{p}-values obtained from statistical tests for random and pseudorandom number generators for cryptographic applications~\cite{NIST2010}
    \item statistical properties of the bitstream hexadecimal representation
\end{itemize}
and we keep the protocol label associated to the bitstream itself.
While the binary representation of the \textit{N} bytes is meant to label the protocol of each packet under analysis, the sequential features allow to understand if the packet content is either compressed or encrypted.

After feature extraction, the raw dataset constituted by streams of bits and their corresponding labels is properly sanitized.
Indeed, it is easily possible to lightly label the network packets belonging to a connection by simply looking either at the ports or at the connection metadata. 
However, this labeling may be prone to errors since it either does not take into account potential custom configurations of services (e.g., SMB protocol operating on a port different from $445$) or intentional misuse of specific protocols by attackers (as in the case of tunneling).
Moreover, cleartext protocols may transfer packets containing compressed data, whose presence could compromise the correct identification of the correct network protocol.
Hence, it is paramount to have a refined and clean dataset to let models perform at their best.
During our experimental evaluations, we have found out that the accuracy of the trained models, after refining the raw dataset, has significantly increased: $7\%$ for the ANN and $20\%$ for compression/encryption detector.
To achieve this performance boost, we have specially implemented an input sanitization module, shown in Figure~\ref{fig:san_mod}.
In this module, we combine unsupervised and supervised support vector machines (SVMs) to clean the raw dataset: first, for each network protocol, we train a one-class SVM both on cleartext and encrypted protocols, in order to filter out outliers from the raw dataset.
As an example, in protocols like HTTP and SMB, requests and responses may contain either the content of (compressed) files or other types of information that are not strictly correlated with the specific protocol communication patterns.
Thus, in order to exclude these outliers, we build one-class SVMs, one for each different protocol, whose hyperparameters are properly tuned on the raw labeled dataset.
Trained models are then applied to identify outliers and remove them from the raw dataset. This refined dataset is then used to train a SVM by applying a one-vs-all classification for detecting packets which are either compressed or encrypted.
This single classifier is applied to remove both compressed and encrypted packets from cleartext protocols.
It is worth mentioning that, in proxied environments, encrypted packets may be present in connections labeled as HTTP: indeed, in these scenarios, also secure communications pass through the proxy, even if these connections are erroneously labeled as HTTP.

This sanitized dataset is then split in training and validation sets to essentially build two different models: (\textit{i}) an artificial neural network (ANN) able to classify cleartext protocols (e.g., DNS) and (\textit{ii}) a SVM that is a compression/encryption detector for identifying, respectively, compressed and encrypted packets.
As later shown in Section~\ref{sec:eval}, after construction, the training set is considerably unbalanced towards secure protocols. 
For this reason, we apply SMOTE data augmentation technique~\cite{SMOTE2002/1622407.1622416} to increase the samples of those protocols belonging to minority classes.
During the test phase, performed light labeling based on connection's destination port is not taken into account and the resulting bitstreams are grouped by connection. 
Each packet is given in input to a trained ANN (whose training process is detailed in Section~\ref{sec:sanitization}) and the analytics both verifies if, in the connection, there are some packets that have been classified with low confidence and more than one protocol is present.
While in this latter case, the co-presence of multiple protocols might signal a possible tunneling attack, when the ANN classifies packets with low confidence, then, the connection could contain either compressed/encrypted packets or packets whose byte sequences differ from the ones usually observed in the network.
To distinguish between these two cases, a more in depth verification is carried out: if the connection is not entirely encrypted, meaning that it is a not a secure communication, the prototype checks if the packets signaled as anomalous (i.e., with low confidence) by the ANN are either encrypted or belongs to another protocol.
If either encryption or compression is detected, the anomaly is notified to security analysts.
On the other hand, if the entire connection is encrypted, it is collected and stored in a database, periodically accessed in order to retrieve data and metadata about X.509, SSL, and TLS exchanged protocol messages in order to be analyzed by the analytics described in~\cite{DBLP:conf/ssci/UcciSBZ21}.

\subsection{Feature extraction}
\label{sec:features}

As discussed in Section~\ref{sec:approach}, sequential features allow to understand if the content of a network packet is either compressed or encrypted.
We rely on a statistical package developed by the Information Technology Laboratory at the National Institute of Standards and Technology, containing a set of $15$ tests that measure the randomness of a binary sequence~\cite{NIST2010}.
These tests have been designed to provide a first step towards the decision whether or not a generated binary sequence can be used in cryptographic applications, namely if the sequence appears to be randomly generated.
In other words, each new bit of the sequence should be unpredictable.
From a statistical point of view, each test verifies if the sequence being under analysis is random.
This null hypothesis can be either rejected or accepted depending on the statistic value on the data exceeding or not a specific value -- called critical value -- that is typically far in the tails of a distribution of reference.
Test reference distributions used in the NIST tests are the standard normal and the $\chi^2$ distributions.
Even if the statistical package contains $15$ tests, we use only $5$ of them, because the length \textit{N} of the binary sequence we test does not meet the corresponding input size recommendation in~\cite{NIST2010}.
To each sequence we apply the following tests: frequency within a block, longest-run-of-ones in a block, serial test, approximate entropy and cumulative sums.
In addition, in our experimental evaluations, we extract some statistical properties and compute the Shannon entropy metrics~\cite{Shannon1948} that, combined with the previously mentioned tests, have shown to improve the overall accuracy of the classification.
As statistical properties, the following features are extracted from the corresponding hexadecimal representation $h$ of a bitstream of \textit{N} bytes:
\begin{itemize}
    \item number of different alphanumeric characters in $h$ normalized over $h$ length
    \item number of different letters in $h$ normalized over $h$ length
    \item longest consecutive sequence of the same character in $h$ normalized over $h$ length
\end{itemize}

\subsection{Input sanitization}
\label{sec:sanitization}

For accurately training machine learning models, the training set should be as much "clean" as possible.
In Section~\ref{sec:approach}, we have already discussed how labeling based on connection metadata could be error prone either due to potential custom configurations of services, intentional misuse of specific protocols by attackers, or network protocols encapsulating compressed data.
In addition, during our experimental evaluations, we have observed that in some cases the employed traffic analyzer can assign an empty label or multiple labels to a single network packet.
While in the first case, bitstreams with empty labels can be easily discarded for the training phase, in the presence of multi-labels is possible to assign a unique correct label if, among the labels, there exist a protocol that is monitored by the prototype itself.
As an example, if the assigned labels are NTLM, GSSAPI, SMB, and DCE\_RPC the resulting label is SMB.
For these reasons, the very first step of the sanitization module is to correct the multi-labels associated to bitstreams and discard the empty ones.
Then, we train an ensamble of one-class SVMs, one for each protocol (see Figure~\ref{fig:san_mod}): each different classifier is properly tuned to filter out outliers from the raw dataset. As stated in Section~\ref{sec:approach}, HTTP and SMB requests or responses may contain either the content of (compressed) files or other types of information that are not strictly correlated with the specific protocol communication patterns.
Trained models are then applied to identify these kind of network packets and remove them from the raw dataset.
This preprocessed dataset is used to train a supervised support vector machine, called compression/encryption detector, by applying a one-vs-all classification for detecting packets which are either compressed or encrypted.
It is worth noting that all these models are still inaccurate because they are trained on a "dirty" dataset.
Hence, to further increase the quality of the labels and obtain the final training set, the compression/encryption detector is fed with cleartext bitstreams to remove possible compressed/encrypted packets from cleartext protocols, as in the case of proxied environments.
The result of this sanization process is a dataset which allows to train and validate two accurate models: an artificial neural network for cleartext protocols and a SVM for compressed and encrypted traffic.

\subsection{Anomaly detection}
\label{sec:anomaly}

During the test phase (see Figure~\ref{fig:proto_flow}), bitstreams are analyzed by the trained ANN.
In turn, the ANN flags three different cases as potential tunneling attacks and alerts security analysts when these cases occur: (\textit{i}) the high confidence detection of more than one protocol in the same connection, (\textit{ii}) the low confidence detection of one protocol for all the packets in the same connection, and (\textit{iii}) the labeling, both with high and low confidence, of one or more protocols for the packets belonging to the same connection (as in the case of secure protocols over DNS).
As later specified in Section~\ref{sec:eval}, in the ANN, the high/low confidence threshold $c$ can be dinamically set.
In any case, the detection of encrypted packets into a cleartext connection generates alert notifications enriched with the information about the presence of encrypted protocol messages.
Possibly, notified alerts can be filtered whitelisting source and/or destination IPs to reduce the false positives caused by well-known machines.

Hence, if some packets of the connection are classified with low confidence, the corresponding bitstream's sequential features (refer to Section~\ref{sec:features}) are given in input to the compression/encryption detector.
If all the packets contained in the connection are encrypted, then the connection and its corresponding metadata are given in input to the SSL/TLS analytics for further scrutiny~\cite{DBLP:conf/ssci/UcciSBZ21}.
On the contrary, if the connection contains some compressed/encrypted packets or none of them, depending on the protocol, the connection is considered anamalous. Indeed, it is worth noting that not in all cases the combination of two different protocols is a signal of an occurring attack: as already discussed, SMB and HTTP connections can contain protocol-specific messages along with compressed data; however, DNS messages interleaved with other protocols are highly suspicious.
\section{Experimental evaluation}
\label{sec:eval}

\begin{table}[tp]
    \begin{minipage}[t]{.48\linewidth}
    \centering
    \captionof{figure}{Packet distribution for each network protocol, before and after balancing.} 
       
        \label{fig:train_proto_distribution}
         \vspace*{5.9mm}
        \includegraphics[scale=0.42]{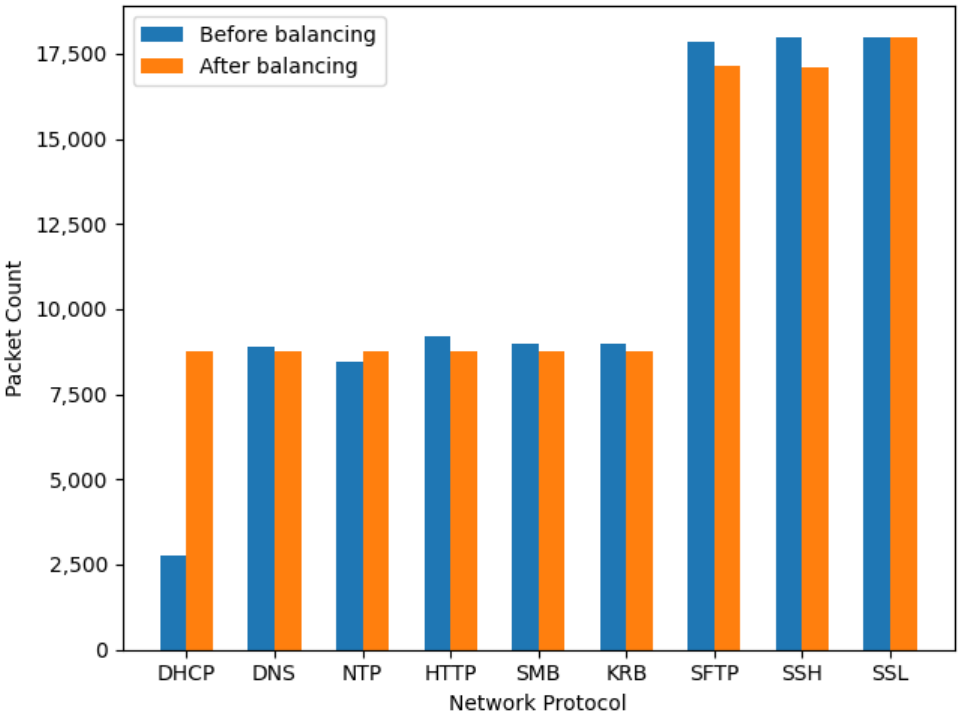} 

    \end{minipage}\hfill
    \quad
    \begin{minipage}[t]{.48\linewidth}
        \centering \vspace{0.19cm}
        \caption{Benign test set summary.}
        \label{tab:ben_test_set}
        \scriptsize
        \begin{tabularx}{\textwidth}{| p{3,38cm} | * {6}{Y|}}
        \hline
        \textit{Statistics} & \multicolumn{1}{| c |}{ \centering \textit{Count [(\%)]}} 
            \\\hline\hline
                DNS packets                   & $30,669$ (1.10\%)   \\ \hline
                SMB packets                   & $65,944$ (2.35\%)    \\ \hline
                HTTP packets                   & $262$ (0.01\%)    \\ \hline
                NTP packets                   & $46$ (0.002\%)    \\ \hline
                DHCP packets                   & $20$ (0.001\%)    \\ \hline
                KRB packets                   & $741$ (0.03\%)    \\ \hline
                SFTP packets                  & $69,158$ (2.46\%)   \\ \hline
                Not labeled packets               & $61,552$ (2.20\%)   \\ \hline
                SSL packets                   &$2,571,608$(91.84\%) \\ \hline	
                Distinct connections          & 51,459      \\ \hline
                Distinct source machines      & 758      \\ \hline
                Distinct dest. machines & 1,566      \\ \hline
        \end{tabularx}
    \end{minipage}\hfill
\end{table}

The proposed prototype and the experimental evaluations have been, respectively, implemented and performed in Python.
The size $N$ we have chosen for the byte sequences, extracted from network packets, is $52$ bytes.
More in detail, we retrieve the first $64$ bytes of the payload of each TCP/UDP packet, from which we remove the first $12$B: indeed, a preliminary evaluation has shown that these first bytes had a very low variance in their binary representation among different packets of the same protocol.
The specific selection of the byte sequence to extract has improved the accuracy of the trained neural network, increasing its anomaly detection capabilities.

For the experimental evaluation of the proposed prototype, we collected both benign and malicious datasets.
The benign communication dataset contains a subset of legitimate traffic observed in a real corporate network during a period of about $2$ days.
From this initial dataset, we sample connections to start building the models' training sets and the dataset that will be used for testing.
Figure~\ref{fig:train_proto_distribution} summarizes general statistics about the collected training set in terms of packets, before and after sanitization. Next to it, Table~\ref{tab:ben_test_set} reports how the test set of legitimate network traffic is characterized.
The sanitization process makes the training set, which is obviously unbalanced towards encrypted protocols, balanced: indeed, after sanitization, the number of packets belonging to respectively cleartext and secure protocols is almost even.
It is worth noting that the balanced training set for the ANN, containing DHCP, DNS, NTP, HTTP and SMB packets, comprises also data belonging to KRB network protocol (i.e., encrypted): our experimental evaluations have shown that during the test phase, the neural network performs better when it is trained also with encrypted byte sequences.
\begin{table}[h!]
	\caption{Support vector machine hyperparameter settings.}
	\label{tab:san_hyper}
	\small
	\centering
	\begin{tabular}{| l | c | c | c | c | c | c |}
		\hline
		
		\multicolumn{1}{|c|}{\textit{Model}} & \multicolumn{1}{|c|}{\textit{Kernel}} & \multicolumn{1}{|c|}{\textit{$\gamma$}} & \multicolumn{1}{|c|}{\textit{$\nu$}} & \multicolumn{1}{|c|}{\textit{$t$}} & \multicolumn{1}{|c|}{\textit{$C$}}
		\\\hline\hline
		DHCP one-class SVM & RBF & $0.7$    & $0.03$   & 0.77 & $-$\\ \hline
		DNS one-class SVM  & RBF & $0.7$    & $0.03$   & 0.77 & $-$\\ \hline
		NTP one-class SVM  & RBF & $0.03$   & $0.1$    & 0.92 & $-$\\ \hline
		HTTP one-class SVM & RBF & $0.08$   & $0.07$   & 0.91 & $-$\\ \hline
		SMB one-class SVM  & RBF & $0.06$   & $0.08$   & 0.77 & $-$\\ \hline
		KRB one-class SVM  & RBF & $0.04$   & $0.05$   & 0.97 & $-$\\ \hline
		SFTP one-class SVM & RBF & $0.7$    & $0.05$   & 0.97 & $-$\\ \hline
		SSH one-class SVM  & RBF & $0.7$    & $0.05$   & 0.97 & $-$\\ \hline
		SSL one-class SVM  & RBF & $0.0001$ & $0.0028$ & 0.97 & $-$\\ \hline
		Compression/encryption detector & RBF & $0.01$ & $-$ & $-$ & $100$ \\ \hline
	\end{tabular}
\end{table}
As ANN, we use a Keras\footnote{Keras library: https://keras.io/} sequential model with $3$ hidden layers.
The input layer accepts $416$ bits (i.e., 52B) and the output layer consists of $6$ neurons, one for each cleartext protocol and KRB.
Regarding SVMs, we rely on the open-source library scikit-learn\footnote{Scikit-learn library: https://scikit-learn.org/stable/index.html}.
For completeness, we report in Table~\ref{tab:san_hyper} the hyperparameters we have used to train the different SVMs in the sanitization module and, in addition, the hyperparameters we obtained by tuning the compression/encryption detector in the validation phase.
It is worth mentioning that the parameter $t$, in Table~\ref{tab:san_hyper}, is used for each protocol one-class SVM as a threshold to filter only those outliers which have a Shannon entropy greater than $t$.
The intuition behind this filtering is that byte sequences having high entropy do not specifically belong to cleartext protocol communications and, thus, they have to be discarded from the training set.

On the other hand, malicious datasets are constituted by packet captures (PCAPs) shared by~\cite{Berg2019}, \cite{CICBell2021}, and~\cite{Iodine2023}.
The former dataset contains $3$ different types of DNS tunnels generated in a controlled environment, whose size are approximately $750$MB each.
Tunneled data contains respectively SFTP, SSH, and Telnet malicious protocol messages. Each sample is made up of one single connection containing millions of DNS packets. It is reasonable to note that such connections would either easily stand out to security analysts or be simply detectable through well-known statistical approaches (e.g., outlier detection). Subsequently, as stated in Section~\ref{sec:approach}, our approach groups data by connection and, therefore, a single malicious packet is enough to flag the entire connection as anomalous. For the above reasons, we have decided to split each sample in $n$ different connections, composed by approximately $5,000$ DNS packets each. The size of the split, reported in Table~\ref{tab:tunnels}, has been chosen according to the size of the connections monitored in the controlled environment.
The second malicious dataset, instead, was born by the collaboration between the Bell Canada company's Cyber Threat Intelligence group and the Canadian Institute for Cybersecurity.
In this dataset, we take only into account DNS packets that, in their payloads, contain exfiltrations of various types of files and we discard legitimate traffic.
Moreover, it is worth mentioning that all the packets contained in~\cite{CICBell2021} have been truncated at capture time to $96$B; this has required a slightly different approach to test these samples that will be discussed later in this Section.
Finally, \cite{Iodine2023} is a single packet capture to test detection and alerting capabilities of Packetbeat\footnote{Elastic Packetbeat: https://www.elastic.co/beats/packetbeat}, Elastic's network packet analyzer.
Malicious packet captures have been injected into the network security platform in order to be processed and analyzed as ordinary traffic.
Table~\ref{tab:tunnels} reports a summary of the malicious assembled datasets: for each PCAP, we list the number of packets in the capture and which ones of these packets have been successfully processed by the platform's network analyzer (i.e., those packets whose size is greater or equal than $64$B); 
in addition, Table~\ref{tab:tunnels} depicts the number of connections in the PCAP and how many of them have been identified as protocol tunneling attacks (i.e., true positives $TP$).
Finally, the true positive rate $TPR$ of the proposed detector is reported for each packet capture.
Analogously, Table~\ref{tab:benign_test_eval} reports the same information contained in Table~\ref{tab:tunnels}, but with reference to the test set described in Table~\ref{tab:ben_test_set}.
\begin{table}[h!]
	\caption{Malicious test set summary.}
	\label{tab:tunnels}
	\footnotesize
	\centering
	\begin{tabular}{| l | c | c | c | c | c | c |}
		\hline
		\multirow{2}{*}{\hspace*{1cm}\textit{Tunnel type}} & \textit{No. of PCAP} & \textit{No. of processed} &  \multirow{2}{*}{\textit{No. of connections}} & \multirow{2}{*}{$TP$} & \multirow{2}{*}{$TPR (\%)$} \\            
		&  \textit{packets} & \textit{PCAP packets} & & &
		\\\hline\hline
		Telnet over DNS tunnel~\cite{Berg2019} & $2.4$M & $2.2$M & $457$ & $457$ & $100\%$\\ \hline
		SFTP over DNS tunnel~\cite{Berg2019} & $2$M & $1$M & $209$ & $209$ & $100\%$\\ \hline
		SSH over DNS tunnel~\cite{Berg2019}  & $2.8$M & $2.7$M & $545$ & $545$ & $100\%$\\ 
		\hline
		\hline
		Light file exfiltration~\cite{CICBell2021} & $187,500$ & $102,000$ & $7,617$ &	$7,361$ & $96.6\%$\\
		\hline
		Heavy file exfiltration~\cite{CICBell2021} & $1.34$M & $765,000$ & $43,964$ & $42,441$ & $96.5\%$ \\ 
		\hline
		\hline
		Data exfiltration over Iodine & \multirow{2}{*}{$438$} & \multirow{2}{*}{$247$} & \multirow{2}{*}{$1$} & \multirow{2}{*}{$1$} & \multirow{2}{*}{$100\%$}\\ 
		DNS tunnel~\cite{Iodine2023} & & & & &\\
		\hline
	\end{tabular}
\end{table}
Being legitimate traffic, the last two columns reports the connections mistakenly classified as tunnels (i.e., false positives $FP$) and the false positive rate $FPR$.
The results of the evaluation, reported in Table~\ref{tab:tunnels} and~\ref{tab:benign_test_eval}, show a false positive rate and a true positive rate, respectively, equal to $5.8\%$ and $96.6\%$.
The overall accuracy of the proposed prototype is $97.1\%$, while the resulting F$1$-score is $95.6\%$.

We conclude this section by discussing how we slightly modified the proposed approach, used in the other datasets, to be compliant with~\cite{CICBell2021}.
Indeed, the DNS packets contained in this dataset have been truncated during traffic acquisition, resulting in byte sequences that do not have the same length.
In order to solve this dataset generation problem, we reduced all the DNS packets to a common length of $44$B, discarding the shorter byte sequences and trimming the longer ones.
The result of the filtering operation is clearly shown in Table~\ref{tab:tunnels}, where the number of processed PCAP packets is more than 54\% less than the ones received in input by the traffic analyzer.

\begin{table}[tp]
	\caption{Benign test set summary.}
	\label{tab:benign_test_eval}
	\small
	\centering
	\begin{tabular}{| c | c | c | c | c | c | c |}
		\hline
		\multirow{2}{*}{\textit{Dataset}} & \textit{No. of PCAP} & \textit{No. of processed} &  \multirow{2}{*}{\textit{No. of connections}} & \multirow{2}{*}{$FP$} & \multirow{2}{*}{$FPR (\%)$} \\            
		& \textit{packets} & \textit{PCAP packets} & & &
		\\\hline\hline
		Legitimate traffic & $5.4$M & $2.8$M & $51,459$ & $2,966$ & $5.8\%$\\ \hline
		
	\end{tabular}
\end{table}
Being the bitstream lengths different from the datasets~\cite{Berg2019} and~\cite{Iodine2023}, we have retrained our ANN to be fed with $44$B sequences.
On the contrary, we have maintained for this evaluation the same hyperparameters for the different SVMs, reported in Table~\ref{tab:san_hyper}, and the same threshold $c$, used in the other experiments.
In particular, for all our experimental evaluations, we set $c$ to $0.999999$ in order to maximize the algorithm sensitivity and to compensate for the lesser information provided by the processing of~\cite{CICBell2021}.
This explains why, in the experimental evaluations, we were not able to achieve a very low false positive rate, as shown in Table~\ref{tab:benign_test_eval}.
However, in context where a high number of false positives could be detrimental, $c$ can be tuned to obtain a 0.5\% false positive rate or lesser without losing accuracy on protocol tunneling attacks.

\section{Conclusion}
\label{sec:conclusion}

In this paper, we proposed a software prototype for detecting protocol tunneling attacks in a monitored network. 
Relying on a combination of machine learning and deep learning techniques, the proposed solution identifies anomalous connections that deviate from the ones usually established in the network.
Since machine learning models are built based only on legitimate traffic, the proposed solution is therefore able to deal with zero-day attacks, because malicious traffic is not required for the learning phase.
The prototype has been evaluated both on malicious and benign datasets: results show a very high accuracy in detecting malicious samples and a low false positive rate on legitimate traffic.

As future work, we plan to optimize the algorithm through  a deeper analysis on how the choice of bytestream length affects the computational time, in order to find a value which guarantees the best trade-off between efficiency and accuracy. 
Indeed, in this work, we mainly focused on accuracy.
Secondly, we envision that the engineered prototype will be integrated into a streaming architecture, where new data are analyzed by the proposed prototype as soon as they are collected to provide the fastest possible response.
In parallel, the models of the protocol tunneling detector are periodically retrained to keep them up to data with possible deviations from the usual behaviour of the monitored network.
Finally, in Section~\ref{sec:anomaly} we outlined the usage of an IP whitelisting filter.
Once in production, the prototype can be easily extended with other security- analyst-defined whitelists as, for example, domain or autonomous system whitelists. 
This will allow the analysts to apply domain-specific knowledge of the monitored network to the protocol tunneling detector, further reducing potential false positives and improving overall performance.

\vspace{6pt} 

\begin{adjustwidth}{-\extralength}{0cm}

\label{sec:bib}
\reftitle{References}
\bibliography{references}

\begin{thebibliography}{999}

\bibitem[ENISA()]{ETL2022}
ENISA Threat Landscape 2022.
\newblock [{O}nline]. {A}vailable:
  \url{https://www.enisa.europa.eu/publications/enisa-threat-landscape-2022}.
  {A}ccessed 06 Feb. 2023.

\bibitem[IBM()]{IBM2022}
Cost of a data breach 2022. A million-dollar race to detect and respond.
\newblock [{O}nline]. {A}vailable:
  \url{{https://www.ibm.com/reports/data-breach}}. {A}ccessed 06 Feb. 2023.

\bibitem[Center for Internet Security()]{Solarwinds2020}
The SolarWinds Cyber-Attack: What You Need to Know.
\newblock [{O}nline]. {A}vailable:
  \url{{https://www.cisecurity.org/solarwinds}}. {A}ccessed 06 Feb. 2023.

\bibitem[Susan Moore()]{Gartner2022}
7 Top Trends in Cybersecurity for 2022.
\newblock [{O}nline]. {A}vailable:
  \url{{https://www.gartner.com/en/articles/7-top-trends-in-cybersecurity-for-2022}}.
  {A}ccessed 06 Feb. 2023.

\bibitem[MITRE()]{TunnelMITRE2020}
Protocol Tunneling.
\newblock [{O}nline]. {A}vailable:
  \url{{https://attack.mitre.org/\\techniques/T1572/}}. {A}ccessed 06 Feb.
  2023.

\bibitem[ENISA()]{ENISAETA2020}
Encrypted Traffic Analysis.
\newblock [{O}nline]. {A}vailable:
  \url{{https://www.enisa.europa.eu/publications/encrypted-traffic-analysis}}.
  {A}ccessed 06 Feb. 2023.

\bibitem[Bisio \em{et~al.}(2017)Bisio, Saeli, Lombardo, Bernardi, Perotti, and
  Massa]{bisio2017real}
Bisio, F.; Saeli, S.; Lombardo, P.; Bernardi, D.; Perotti, A.; Massa, D.
\newblock Real-time behavioral DGA detection through machine learning.
\newblock In Proceedings of the International Carnahan Conference on Security
  Technology (ICCST). IEEE,  2017, pp. 1--6.
\newblock {\url{https://doi.org/10.1109/ccst.2017.8167790}}.

\bibitem[Lombardo \em{et~al.}(2018)Lombardo, Saeli, Bisio, Bernardi, and
  Massa]{lombardo2018fast}
Lombardo, P.; Saeli, S.; Bisio, F.; Bernardi, D.; Massa, D.
\newblock Fast Flux Service Network Detection via Data Mining on Passive DNS
  Traffic.
\newblock In Proceedings of the International Conference on Information
  Security. Springer,  2018, pp. 463--480.
\newblock {\url{https://doi.org/10.1007/978-3-319-99136-8\_25}}.

\bibitem[Saeli \em{et~al.}(2020)Saeli, Bisio, Lombardo, and Massa]{Saeli2020}
Saeli, S.; Bisio, F.; Lombardo, P.; Massa, D.
\newblock DNS Covert Channel Detection via Behavioral Analysis: a Machine
  Learning Approach.
\newblock In Proceedings of the International Conference on Malicious and
  Unwanted Software (MALWARE),  2020, pp. 46--55.

\bibitem[Ucci \em{et~al.}(2021)Ucci, Sobrero, Bisio, and
  Zorzino]{DBLP:conf/ssci/UcciSBZ21}
Ucci, D.; Sobrero, F.; Bisio, F.; Zorzino, M.
\newblock Near-real-time Anomaly Detection in Encrypted Traffic using Machine
  Learning Techniques.
\newblock In Proceedings of the {IEEE} Symposium Series on Computational
  Intelligence, {SSCI} 2021, Orlando, FL, USA, December 5-7, 2021. {IEEE},
  2021, pp. 1--8.
\newblock {\url{https://doi.org/10.1109/SSCI50451.2021.9659955}}.

\bibitem[Felt \em{et~al.}(2017)Felt, Barnes, King, Palmer, Bentzel, and
  Tabriz]{Felt2017}
Felt, A.P.; Barnes, R.; King, A.; Palmer, C.; Bentzel, C.; Tabriz, P.
\newblock Measuring HTTPS Adoption on the Web.
\newblock In Proceedings of the Proceedings of the 26th USENIX Conference on
  Security Symposium,  2017, p. 1323–1338.

\bibitem[Chad Skipper()]{VMware2020}
The Relevance of Network Security in an Encrypted World.
\newblock [{O}nline]. {A}vailable:
  \url{{https://blogs.vmware.com/networkvirtualization/2020/09/network-security-encrypted.html/}}.
  {A}ccessed 06 Feb. 2023.

\bibitem[Decipher()]{Decipher2019}
Encryption, Privacy in the Internet Trends Report.
\newblock [{O}nline]. {A}vailable:
  \url{{https://duo.com/decipher/encryption-privacy-in-the-internet-trends-report}}.
  {A}ccessed 06 Feb. 2023.

\bibitem[Nirav Shah()]{Fortinet2020}
Keeping Up With the Performance Demands of Encrypted Web Traffic.
\newblock [{O}nline]. {A}vailable:
  \url{{https://www.fortinet.com/blog/industry-trends/keeping-up-with-performance-demands-of-encrypted-web-traffic}}.
  {A}ccessed 06 Feb. 2023.

\bibitem[Google()]{GoogleChrome2020}
Google Transparency Report: HTTPS encryption on the web.
\newblock [{O}nline]. {A}vailable:
  \url{{https://transparencyreport.google.com/https/overview?hl=en}}.
  {A}ccessed 06 Feb. 2023.

\bibitem[Cisco()]{Cisco2019}
Cisco Encrypted Traffic Analytics.
\newblock [{O}nline]. {A}vailable:
  \url{{https://www.cisco.com/c/en/us/solutions/collateral/enterprise-networks/enterprise-network-security/nb-09-encrytd-traf-anlytcs-wp-cte-en.pdf}}.
  {A}ccessed 06 Feb. 2023.

\bibitem[ENISA()]{ENISA2020}
ENISA Threat Landscape - Malware.
\newblock [{O}nline]. {A}vailable:
  \url{{https://www.enisa.europa.eu/publications/malware/at\_download/fullReport}}.
  {A}ccessed 06 Feb. 2023.

\bibitem[Korolov(2012)]{korolov2012cyber}
Korolov, M.
\newblock Cyber Security Review.
\newblock {\em Treasury \& Risk} {\bf 2012}.

\bibitem[Taylor \em{et~al.}(2014)Taylor, Fritsch, and
  Liederbach]{taylor2014cyber}
Taylor, R.W.; Fritsch, E.J.; Liederbach, J.
\newblock {\em Digital crime and digital terrorism}; Prentice Hall Press,
  2014.

\bibitem[Yadav and Mallari(2016)]{yadav2016cyber_sec}
Yadav, T.; Mallari, R.A.
\newblock Technical aspects of cyber kill chain.
\newblock {\em arXiv preprint arXiv:1606.03184} {\bf 2016}.

\bibitem[David Hoelzer()]{SANS2021}
Applying Machine Learning to Network Anomalies.
\newblock [{O}nline]. {A}vailable:
  \url{{https://www.youtube.com/watch?v=qOfgNd-qijI}}. {A}ccessed 06 Feb. 2023.

\bibitem[Wang \em{et~al.}(2021)Wang, Zhou, Liao, Zheng, Hu, and
  Zhang]{wang2021comprehensive}
Wang, Y.; Zhou, A.; Liao, S.; Zheng, R.; Hu, R.; Zhang, L.
\newblock A comprehensive survey on DNS tunnel detection.
\newblock {\em Computer Networks} {\bf 2021}, {\em 197},~108322.

\bibitem[Palau \em{et~al.}(2020)Palau, Catania, Guerra, Garcia, and
  Rigaki]{palau2020dns}
Palau, F.; Catania, C.; Guerra, J.; Garcia, S.; Rigaki, M.
\newblock DNS tunneling: A deep learning based lexicographical detection
  approach.
\newblock {\em arXiv preprint arXiv:2006.06122} {\bf 2020}.

\bibitem[Zhang \em{et~al.}(2019)Zhang, Yang, Yu, and Ma]{zhang2019dns}
Zhang, J.; Yang, L.; Yu, S.; Ma, J.
\newblock A DNS tunneling detection method based on deep learning models to
  prevent data exfiltration.
\newblock In Proceedings of the Network and System Security: 13th International
  Conference, NSS 2019, Sapporo, Japan, December 15--18, 2019, Proceedings 13.
  Springer,  2019, pp. 520--535.

\bibitem[Ahmed \em{et~al.}(2019)Ahmed, Gharakheili, Raza, Russell, and
  Sivaraman]{ahmed2019real}
Ahmed, J.; Gharakheili, H.H.; Raza, Q.; Russell, C.; Sivaraman, V.
\newblock Real-time detection of DNS exfiltration and tunneling from enterprise
  networks.
\newblock In Proceedings of the 2019 IFIP/IEEE Symposium on Integrated Network
  and Service Management (IM). IEEE,  2019, pp. 649--653.

\bibitem[Rajendran \em{et~al.}(2020)Rajendran et~al.]{rajendran2020dns}
Rajendran, B.;  et~al.
\newblock DNS amplification \& DNS tunneling attacks simulation, detection and
  mitigation approaches.
\newblock In Proceedings of the 2020 International Conference on Inventive
  Computation Technologies (ICICT). IEEE,  2020, pp. 230--236.

\bibitem[Vapnik(2013)]{vapnik2013nature}
Vapnik, V.
\newblock {\em The nature of statistical learning theory}; Springer science \&
  business media,  2013.

\bibitem[Swersky \em{et~al.}(2016)Swersky, Marques, Sander, Campello, and
  Zimek]{swersky2016evaluation}
Swersky, L.; Marques, H.O.; Sander, J.; Campello, R.J.; Zimek, A.
\newblock On the evaluation of outlier detection and one-class classification
  methods.
\newblock In Proceedings of the 2016 IEEE International Conference on Data
  Science and Advanced Analytics (DSAA). IEEE,  2016, pp. 1--10.
\newblock {\url{https://doi.org/10.1109/dsaa.2016.8}}.

\bibitem[Abiodun \em{et~al.}(2018)Abiodun, Jantan, Omolara, Dada, Mohamed, and
  Arshad]{abiodun2018state}
Abiodun, O.I.; Jantan, A.; Omolara, A.E.; Dada, K.V.; Mohamed, N.A.; Arshad, H.
\newblock State-of-the-art in artificial neural network applications: A survey.
\newblock {\em Heliyon} {\bf 2018}, {\em 4},~e00938.

\bibitem[NIST()]{NIST2010}
A Statistical Test Suite for Random and Pseudorandom Number Generators for
  Cryptographic Applications.
\newblock [{O}nline]. {A}vailable:
  \url{https://nvlpubs.nist.gov/nistpubs/legacy/sp/nistspecialpublication800-22r1a.pdf}.
  {A}ccessed 06 Feb. 2023.

\bibitem[Chawla \em{et~al.}(2002)Chawla, Bowyer, Hall, and
  Kegelmeyer]{SMOTE2002/1622407.1622416}
Chawla, N.V.; Bowyer, K.W.; Hall, L.O.; Kegelmeyer, W.P.
\newblock SMOTE: Synthetic Minority over-Sampling Technique.
\newblock {\em J. Artif. Int. Res.} {\bf 2002}, {\em 16},~321–357.

\bibitem[Shannon(1948)]{Shannon1948}
Shannon, C.E.
\newblock A Mathematical Theory of Communication.
\newblock {\em The Bell System Technical Journal} {\bf 1948}, {\em
  27},~379--423.

\bibitem[Berg and Forsberg(2019)]{Berg2019}
Berg, A.; Forsberg, D.
\newblock Identifying DNS-tunneled traffic with predictive models.
\newblock {\em CoRR} {\bf 2019}, {\em abs/1906.11246},
  \href{http://xxx.lanl.gov/abs/1906.11246}{{\normalfont [1906.11246]}}.

\bibitem[Mahdavifar \em{et~al.}(2022)Mahdavifar, Hanafy~Salem, Victor, Razavi,
  Garzon, Hellberg, and Lashkari]{CICBell2021}
Mahdavifar, S.; Hanafy~Salem, A.; Victor, P.; Razavi, A.H.; Garzon, M.;
  Hellberg, N.; Lashkari, A.H.
\newblock Lightweight Hybrid Detection of Data Exfiltration Using DNS Based on
  Machine Learning.
\newblock In Proceedings of the 2021 the 11th International Conference on
  Communication and Network Security; Association for Computing Machinery: New
  York, NY, USA,  2022; ICCNS 2021, p. 80–86.
\newblock {\url{https://doi.org/10.1145/3507509.3507520}}.

\bibitem[Elastic()]{Iodine2023}
Iodine DNS Tunnel.
\newblock [{O}nline]. {A}vailable:
  \url{{https://github.com/elastic/examples/blob/master/Security\%20Analytics/dns_tunnel_detection/dns-tunnel-iodine.pcap}}.
  {A}ccessed 06 Feb. 2023.

\end{thebibliography}

\PublishersNote{}
\end{adjustwidth}
\end{document}